\documentclass[floatfix,onecolumn,pre]{revtex4}
\usepackage[dvips]{graphicx}
\usepackage{amssymb}
\usepackage{amsmath}
\usepackage{latexsym}
\usepackage{epsfig}
\usepackage{bm}
\usepackage{times,psfrag,subfigure} 
\usepackage[dvips]{color}       
\newcommand{\young}{\ensuremath{\theta_{\mathrm{Y}}}}
\newcommand{\cassie}{\ensuremath{\theta_{\mathrm{C}}}}
\newcommand{\anticassie}{\ensuremath{\bar{\theta}_{\mathrm{C}}}}
\DeclareMathOperator{\cn}{cn}
\DeclareMathOperator{\dn}{dn}

\begin{document}

\title{Superhydrophobicity on hairy surfaces}
\author{M. L. Blow}
\author{J. M. Yeomans}
\affiliation{The Rudolf Peierls Centre for Theoretical Physics, Oxford University, 1 Keble Road, Oxford OX1 3NP, England}
\date{\today}


\begin{abstract}
We investigate the wetting properties of surfaces patterned with fine elastic hairs, with an emphasis on identifying 
superhydrophobic states on hydrophilic hairs. We formulate a two dimensional model of a large drop in contact with 
a row of equispaced elastic hairs and, by minimising the free energy of the model, identify the stable and metastable states. 
In particular we concentrate on ‘partially suspended’ states, where the hairs bend to support the drop -- singlet states where all hairs bend in the same direction, and doublet states where neighbouring hairs bend in opposite directions -- and find the limits of stability of these configurations in terms of material contact angle, hair flexibility, and system geometry. The drop can remain suspended in a singlet state at hydrophilic contact angles, but doublets exist only when the hairs are hydrophobic. The system is more likely to evolve into a singlet state if the hairs are inclined at the root. We discuss how, under limited circumstances, the results can be modified to describe an array of hairs in three dimensions. We find that now both singlets and doublets can exhibit superhydrophobic behaviour on hydrophilic hairs. We discuss the limitations of our approach and the directions for future work.
\end{abstract}
\maketitle



\section{Introduction}
\label{section:Introduction}
There is widening interest in understanding how fluid streams and drops interact with micropatterned surfaces. For example chemical patterning can be used to direct streams of fluid, and hydrophobic surfaces which are rough on the micron or nanometre length scale can exhibit superhydrophobicity, characterised by contact angles near $180^{\circ}$ and easy roll-off~\cite{Quere}. A primary motivation to study micropatterned substrates comes from their potential applications as, for example, more efficient water and dew repellent materials, dehumidifiers or low drag surfaces. Inspiration for these developments has come in part from nature where plants, such as the lotus~\cite{Barthlott} and nasturtium, have superhydrophobic leaves, and water-walking insects, like the water strider~\cite{Gao}, have superhydrophobic legs to allow them to interact with the the water in their environment.\\

Many of the natural surfaces that show strong water repellency are covered with tiny hairs. These can be long and flexible as on the leaves of Lady's Mantle or form a dense, spiky array as on the legs of the water strider. There have also been recent advances in the microfabrication of hairy surfaces~\cite{MockForsterEtAl,HsuSigmund}. To fully exploit these possibilities it is important to gain a better theoretical understanding of how fluids interact with hairy surfaces, and here we aim to make progress in this direction. In particular, we ask how the elasticity, wetting angle and incline of an array of hairs determine their effectiveness in supporting a liquid away from the base substrate.\\

A water drop deposited on a partially wetting solid does not spread indefinitely, but remains as a localised spherical cap, making a finite angle with the substrate. This angle is determined by the balance of surface tensions between water, air and solid surface, as given by Young's law~\cite{Young}
\begin{align}
\cos\young=\frac{\gamma_{\mathrm{SA}}-\gamma_{\mathrm{SW}}}{\gamma}\;,     \label{eqn:young}
\end{align}
where $\gamma$, $\gamma_{\mathrm{SA}}$ and $\gamma_{\mathrm{SW}}$ are the surface tensions of water-air, substrate-air and substrate-water contact. $\young$, the Young angle, describes a hydrophilic, hydrophobic or neutrally-wetting surface depending on whether it is below, above or equal to $90^{\circ}$ respectively. Using known materials, the Young angle that can be achieved for water on a smooth substrate is no more than $120^{\circ}$. However some rough surfaces can produce much higher contact angles. This occurs when the drop lies on top of the topographical structure, in the suspended or Cassie-Baxter state~\cite{Cassie}. In the suspended state, the extra air-liquid interface at the base of the drop can result in a very high contact angle, even when the material is only slightly hydrophobic. Moreover contact angle hysteresis is small and the drop rolls over the surface very easily~\cite{Quere}, so the suspended state is superhydrophobic.\\

The drop can also penetrate the interstices between the relief, to lie in contact with all points of the substrate, in the collapsed or Wenzel state~\cite{Wenzel}. The collapsed state boosts the contact angle only modestly, and increases resistance to a drop's motion~\cite{Quere}, so is usually not considered superhydrophobic. Ensuring the feasiblilty and robustness of the suspended state is therefore an important consideration in the design of superhydrophobic surfaces.\\

Facilitating a suspended state for {\it hydrophilic} Young angles would also have important applications. A major weakness of many superhydrophobic surfaces is that they fail to repel liquids other than water, for example hydrocarbons. This is because there are virtually no materials for which $\young$ with respect to these liquids is greater than $90^{\circ}$~\cite{Zisman}. Oily substances will sit in the collapsed state or, worse, be imbibed into the structure. This will despoil the surface, because a water drop will now be in contact with the oil, rather than an air layer. If the surface can be made superoleophobic~\cite{TutejaEtAl}, sustaining a suspended state for Young angles below $90^{\circ}$, then this problem will be solved.\\

Posts on a substrate with a degree of overhang may permit suspension, even when the Young angle is hydrophilic, by pinning the contact line~\cite{LiuEtAl}. Such surfaces have been constructed using mushroom shaped posts~\cite{TutejaEtAl,KimEtAl}. As an alternative to overhanging geometries, we investigate the possibility that flexible posts or hairs on a surface may deform to support water away from the surface, even when they are hydrophilic.\\

The leaves of the Lady's Mantle ({\it Alchemilla Mollis/Vulgaris}) are superhydrophobic~\cite{OttenHerminghaus,BrewerWillis}, with water forming beads on them. Closer inspection shows them to be covered with fine hairs, just visible to the naked eye. Surprisingly, experiments~\cite{OttenHerminghaus} found these hairs to be hydrophilic. Otten and Herminghaus~\cite{OttenHerminghaus} stressed the importance of the flexibility of the hairs in sustaining the suspended state. They proposed that the hairs bunch together to reduce distortion of the interface, and the consequent energy cost of bending the hairs prevents them from being entirely wetted. These arguments were disputed by Berndino et al~\cite{Bernardino}, who cast doubt on both the viability of the mechanism, and the elasticity of the Lady's Mantle hairs.\\

Some arthropods, such as the water spider and the water boatman, can dive underwater for substantial periods. Observations show a shiny appearance to the underside of these animals~\cite{NobleNesbitt}, which is due to a layer of air trapped against the body, called a plastron. The plastron provides the means of respiration~\cite{deRuiterEtAL}, not by acting as an artificial lung (a store of air taken from above water) but as an artificial gill (an interface through which gas exchange with the water may take place)~\cite{Ege}. Thorpe and Crisp~\cite{ThorpeCrispA,ThorpeCrispB,ThorpeCrispC,CrispThorpe,ThorpeCrispD} demonstrated that the air layer is trapped by a bed of hairs, such that the surrounding water is maintained in a suspended, superhydrophobic state. In contrast to the Lady's Mantle, it has been argued that flexibility undermines the water repellency of the plastron of a diving arthropod, with any bunching of hairs leaving gaps through which the water will invade~\cite{ThorpeCrispD,Bush}. Conversely, it has been argued that flexibility of the hairs on an insect's legs can provide directional control as it walks on the pond surface~\cite{Bush}. These examples suggest that hair flexibility modifies the wetting properties of a surface in a rich and nontrivial way.\\

A mathematical model of the static configurations of a drop on a hairy surface can be constructed by treating the drop as a capillary surface and the hairs as elastica. In recent years, there have been a number of such elastocapillary analyses~\cite{Bernardino,CohenMahadeven,BicoRomanEtAl,BoudaoudEtAl,NeukirchEtAl,ParkKim,KimMahadevan,Vella,KwonEtAl}. Of particular relevance to the work we present in this paper are those investigating the interaction of initally flat elastic sheets with an initially flat fluid interface. Neukirch et al~\cite{NeukirchEtAl}, found that a perfectly wetting elastic beam, raised and lowered from a tank of liquid, exhibited hysteretic behaviour, arising from the metastable binding of the beam to the interface. Park and Kim~\cite{ParkKim} similarly showed that the threshold load for a clamped rod to pierce an interface was greater for a flexible rod than for a rigid rod. Kim and Mahadevan~\cite{KimMahadevan} studied the problem of capillary rise between a pair of elastic sheets arguing that, at modest flexibility, the height of capillary rise is increased by the narrowing of the gap, but for higher flexibilities the sheets can seal together, thus preventing further rise. These three studies~\cite{NeukirchEtAl,ParkKim,KimMahadevan} take the capillary length to be comparable to the system size so that gravity is an important influence which sets a decay length for disturbances of the fluid interface. The study we present in this paper differs from these by considering a periodic system with a capillary length much longer than the periodicity. Gravity may therefore be ignored, leaving just a balance between the interfacial and bending energies.\\

The model we treat is an interface supported by a line of flexible hairs. In general, the hairs will lie along the interface, thus alleviating the surface tension of the drop. We obtain an exact analytic solution in two dimensions which allows us to describe the possible drop configurations and obtain quantitative boundaries for the transitions between them.  For vertical hairs, as for rigid posts, a drop placed in the suspended state can remain there only if the hairs are hydrophobic. However, for both hydrophobic and hydrophilic hairs, we identify a partially suspended state where the interface lies between the tops of the hairs and the surface. We show that, once the drop is in the partially suspended configuration it can remain there even for hydrophilic hairs. We discuss whether it is feasible for a drop placed gently on the surface to access the partially suspended state, and show that this is rather natural, if the hairs have a small angle of inclination to the vertical.\\

The layout of this paper is as follows: In Sec.~\ref{section:2Dmodel} we describe the model and characterise the possible stable and metastable interface configurations. In Sec.~\ref{section:freeEnergy2D} we write down the free energy and obtain equations for the profile of the hairs. The stability of partially suspended singlet and doublet states, which have the periodicity or twice the periodicity of the lattice of hairs, are discussed, as a function of the hair flexibility and Young angle, in Secs.~\ref{section:singlets} and~\ref{section:doublets} respectively. We then, in Sec.~\ref{section:latticeBoltzmann}, present numerical results showing how the partially suspended states can be formed. Sec.~\ref{section:2Dincline} considers inclined hairs, and Sec.~\ref{section:LaplacePressure} discusses the effects of finite drop curvature.  In Sec.~\ref{section:3Dmodel} we extend our results to three dimensions. A conclusion summarises the paper and discusses the many directions for further research.

\section{Model for a line of hairs}
\label{section:2Dmodel}
\begin{figure}
\centering
{\epsfig{file=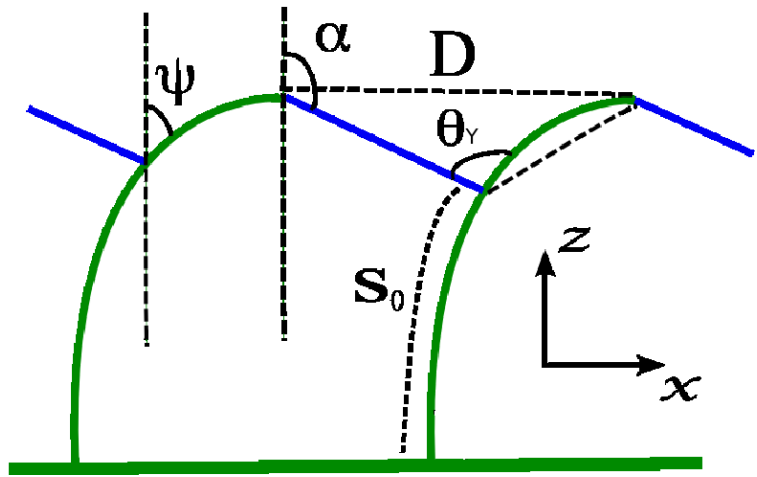,width=100mm}}
\caption{Illustration of the $\mathbb{P}_{1}$ state, showing some of the parameters and variables 
of the system. The hairs are shown as green (lighter grey) and the liquid-air interface as blue (darker grey).}
\label{fig:singlets}
\end{figure}
\begin{figure}
\centering
{\epsfig{file=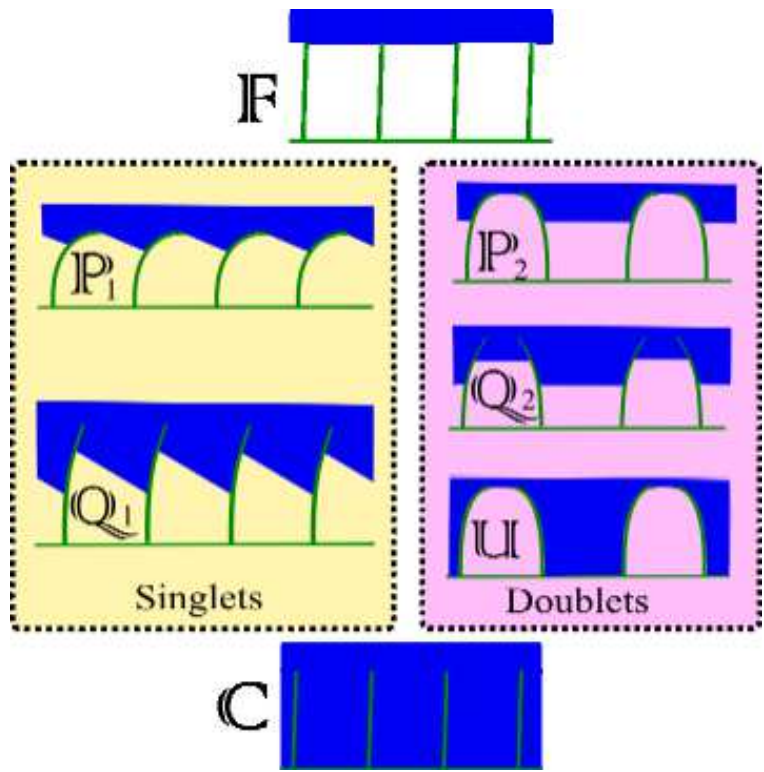,width=100mm}}
\caption{Schematic illustrations of the periodic states of the system. The hairs are shown as green (lighter grey) and the liquid-air interface as blue (darker grey). At the top is the fully suspended state, $\mathbb{F}$, and at the bottom is the collapsed state, $\mathbb{C}$. On the left are the singlets states: partially suspended, $\mathbb{P}_{1}$, and quasi-suspended $\mathbb{Q}_{1}$. On the right are the doublet states: partially suspended, $\mathbb{P}_{2}$, quasi-suspended $\mathbb{Q}_{2}$, and unzipped $\mathbb{U}$. See section \ref{section:2Dmodel} for details.}
\label{fig:2Dstates}
\end{figure}
We consider an infinite row of hairs (or equivalently elastic sheets with no bending along $y$), labelled with integers $n$, attached to a flat substrate,
with regular spacing $D$ along the $x$ direction, and with a fixed inclination $\Omega$ to the $z$
direction (see Fig.~\ref{fig:singlets}). The hairs are inextensible, of length $L$ and width negligible compared to $D$, and have a bending modulus $K$. They are in contact with the base of a liquid drop of surface tension $\gamma$. The surfaces of the hairs and the substrate are smooth and have a contact angle with respect to the liquid of the drop of $\young$. The drop is assumed to be sufficiently small that gravity can be neglected, but large enough that 
Laplace curvature and edge effects are unimportant, $(L,D)\ll R\ll\lambda_{\mathrm{C}}$ where $R$
is the drop radius and $\lambda_{\mathrm{C}}$ is the capillary length. In this regime the interface 
between neighbouring hairs is straight. Note that a hair may contact the interface in different
places on its two sides. This is a consequence of working in two dimensions. We discuss the circumstances under which our results are relevant to hairs in three dimensions in Sec.~\ref{section:3Dmodel}.\\

We introduce a classification scheme for the stable interface configurations, which are illustrated
in Fig.~\ref{fig:2Dstates}. In the rigid limit it is well known that the possible states are the
{\it fully suspended} or Cassie-Baxter state, in which the interface rests on the tips of the
hairs, and the {\it collapsed} or Wenzel state in which the liquid wets the hairs and the base
substrate. We denote these $\mathbb{F}$ and $\mathbb{C}$ respectively. The symmetry of the
interface configurations implies that the hairs in either of these states will be straight for all
values of the hair rigidity. We choose the free energy of the fully suspended state $\mathbb{F}$ to
be zero. Therefore the free energy of the collapsed state, $\mathbb{C}$ is per length $D$ along the surface.
\begin{equation}
\begin{split}
\mathcal{F}_{\mathbb{C}}&=-\gamma D+\left(\gamma_{\mathrm{SW}}-\gamma_{\mathrm{SA}}\right)(D+2L)\\
&=\gamma\left[-D-\cos\young(D+2L)\right]
\end{split}
\end{equation}
where Eqn.~(\ref{eqn:young}) has been used in the second step. \\

Flexible hairs also permit {\it partially suspended} states, where the interface on one side of a 
hair is pinned at its tip, but that on the other side meets the hair part-way up, at an arc length
$s_{0}$ from the base of the hair, say (see Fig.~\ref{fig:singlets}). The simplest partially suspended state, 
which we shall term the {\it singlet} and denote  $\mathbb{P}_{1}$, has periodicity $D$. Each hair
bends in an identical curve, in the same direction, and a section of interface connects the tip of 
hair $n$ to the point $s_{0}$ on hair $n+1$.  A second possibility is a {\it doublet}, $\mathbb{P}_{2}$, 
with periodicity $2D$, where each hair takes the same shape to within a reflection, but 
neighbouring hairs bend in opposite directions. In this state, the section of interface between
hairs $n$ and $n+1$ joins the tips of neighbouring hairs for $n$ even, and the points $s_0$ for $n$
odd. Hence the interface always lies parallel to the substrate, but at two different heights.\\

Note that $\mathbb{F}$ is equivalent to the partially suspended states for $s_{0}= L$. As $s_{0}\to
0$ in the $\mathbb{P}_{1}$ configuration, the interface meets the substrate at an obtuse angle and
the fluid readily wets the surface  together with the dry sides of the hairs resulting in the
collapsed state $\mathbb{C}$. By contrast, as $s_{0}\to 0$
in the $\mathbb{P}_{2}$ configuration the substrate can wet between hairs $n$ and $n+1$, $n$ odd,
but remain suspended for $n$ even, resulting in air pockets. We shall refer to such configurations 
as  {\it unzipped} states, $\mathbb{U}$ (see Fig.~\ref{fig:2Dstates}). Other conceivable states are {\it quasi-suspended} states
$\mathbb{Q}_{1,2}$ where the interface meets a hair part-way along on both sides. We shall show, in
section~\ref{section:singlets}, that such states are not stable here.\\
 
 
\section{The free energy}
\label{section:freeEnergy2D}
In this section we write down a free energy for the partially suspended states $\mathbb{P}_{1,2}$, starting with the case $\Omega=0$. The shape of the hair is defined by the position vector $\mathbf{r}(s)$ where $s$ is the arc length which runs from $0$ at the base to $L$ at the tip. Because $s$ is a measure of arclength, $\dot{\mathbf{r}}(s)$ is the unit tangent to the hair at $s$ and $\ddot{\mathbf{r}}(s)$ is the directed curvature. The free energy of the system, measured for a length $D$ along the surface, and measured relative to the free energy of the fully suspended state, is
\label{section:2Dzero}
\begin{equation}
\mathcal{F}=\int_{0}^{s_{0}}\left[\tfrac{1}{2}K\ddot{\mathbf{r}}^{2}+\tfrac{1}{2}\sigma(s)\left(\dot{\mathbf{r}}^{2}-1\right)\right]ds+\int_{s_{0}}^{L}\left[\tfrac{1}{2}K\ddot{\mathbf{r}}^{2}+\tfrac{1}{2}\sigma(s)\left(\dot{\mathbf{r}}^{2}-1\right)-\gamma\cos\young\right]ds+\gamma(\vert\pmb{\Lambda}\vert-D).   \label{eqn:freeEnergyZeroP}
\end{equation}
The first two terms in each of the integrals correspond to modelling the hairs as inextensible Euler elastica. The first term denote a free energy quadratic in the curvature. (Such a functional form may be justified from the considering the hair to be made of a linear elastic material~\cite{LandauLifschitz}, with $K=EI$, where $E$ is the Young's modulus of the material and $I$ is a geometrical moment of the hair's cross section.) The third term in the second integral is the free energy which results from the liquid wetting the hairs. The remaining term is the liquid-gas interfacial energy where
\begin{equation}
\pmb{\Lambda}=\begin{cases}
D\mathbf{e}_{x}-\left(\mathbf{r}(L)-\mathbf{r}(s_{0})\right)\;,&\text{for}\;\; \mathbb{P}_{1}\;,\\
\left[D-\left(x(L)-x(s_{0})\right)\right]\mathbf{e}_{x}\;,&\text{for} \;\; \mathbb{P}_{2}\;,
\end{cases}
\end{equation}
is the length of interface between successive hairs.\\

We split the minimisation of the free energy into two stages, first applying the calculus of 
variations for fixed $s_{0}$, and then minimising with respect to $s_{0}$. In the first step we consider the variation of the free energy $\mathcal{F}\rightarrow\mathcal{F}+\delta\mathcal{F}$ with respect to infinitesimal perturbations of the curve $\mathbf{r}\rightarrow\mathbf{r}+\delta\mathbf{r}$. If the hair is in a configuration corresponding to an extremum of $\mathcal{F}$ then $\delta\mathcal{F}$ will be zero to first order in all $\delta\mathbf{r}$. However, we must constrain the set of solutions to those where the parameterisation corresponds to $s$ being the arclength along the curve. This is important for two reasons: to ensure the hair has the correct length, and to ensure that $\ddot{\mathbf{r}}$ corresponds to curvature. To do this, we also vary $\sigma(s)$, independently of $\mathbf{r}(s)$. $\sigma(s)$ may be likened to a stress in the hair, resisting extension or compression. Considering variation of $\mathcal{F}$ with respect to $\mathbf{r}(s)$ and $\sigma(s)$,
\begin{equation}
\begin{split}
\delta\mathcal{F}=&\int_{0}^{s_{0}}\left\{K\ddot{\mathbf{r}}\mathbf{.}\delta\ddot{\mathbf{r}}+\sigma
\dot{\mathbf{r}}\mathbf{.}\delta\dot{\mathbf{r}}+\left(\dot{\mathbf{r}}^{2}-1\right)\delta\sigma\right\}ds+\int_{s_{0}}^{L}\left\{K\ddot{\mathbf{r}}\mathbf{.}\delta\ddot{\mathbf{r}}+\sigma\dot{\mathbf{r}}\mathbf{.}\delta\dot{\mathbf{r}}+\left(\dot{\mathbf{r}}^{2}-1\right)\delta\sigma\right\}ds+\gamma\frac{\pmb{\Lambda}.\delta\pmb{\Lambda}}{\lvert\pmb{\Lambda}\rvert}\\
=&\int_{0}^{s_{0}}\left\{\left(K\ddddot{\mathbf{r}}-\frac{d(\sigma\dot{\mathbf{r}})}{ds}\right)\mathbf{.}\delta\mathbf{r}+\left(\dot{\mathbf{r}}^{2}-1\right)\delta\sigma\right\}ds+\left[-K\left(\ddot{\mathbf{r}}^{+}-\ddot{\mathbf{r}}^{-}\right)\mathbf{.}\delta\dot{\mathbf{r}}+\left(K\left(\dddot{\mathbf{r}}^{+}-\dddot{\mathbf{r}}^{-}\right)-\left(\sigma^{+}-\sigma^{-}\right)\dot{\mathbf{r}}+\pmb{\gamma}\right)\mathbf{.}\delta\mathbf{r}\right]_{s_{0}}\\&+\int_{s_{0}}^{L}\left\{\left(K\ddddot{\mathbf{r}}-\frac{d(\sigma\dot{\mathbf{r}})}{ds}\right)\mathbf{.}\delta\mathbf{r}+\left(\dot{\mathbf{r}}^{2}-1\right)\delta\sigma\right\}ds+\left[K\ddot{\mathbf{r}}\mathbf{.}\delta\dot{\mathbf{r}} +\left(-K\dddot{\mathbf{r}}+\sigma\dot{\mathbf{r}}-\pmb{\gamma}\right)\mathbf{.}\delta\mathbf{r}\right]_{L}\;    \label{eqn:freeEnergyVariations}
\end{split}
\end{equation}

where we have expressed the surface tension as a force vector 
$\pmb{\gamma}=\gamma\frac{\pmb{\Lambda}}{\vert\pmb{\Lambda}\vert}$, and used the fact that 
$\mathbf{r}$ and $\dot{\mathbf{r}}$ must be continuous if $\mathcal{F}$ is not to diverge. A 
continuity condition does not automatically apply to other quantities; hence the $\pm$ superscripts 
to denote the direction from which the limit $s\rightarrow s_{0}$ is taken.\\

To find the extremal solution we set $\delta\mathcal{F}=0$. Since $\delta\mathbf{r}$ and $\delta\sigma$ are considered independent and arbitrary functions, the coefficients of each of these in the integrand of (\ref{eqn:freeEnergyVariations}) must be zero for all $s$ for the condition $\delta\mathcal{F}=0$ to hold identically. From the coefficient of $\delta\sigma$,
\begin{equation}
\dot{\mathbf{r}}^{2}-1=0\;,  \label{eqn:tangentCondition}
\end{equation}
or $\lvert\dot{\mathbf{r}}\rvert=1$, thus guaranteeing that $s$ measures arclength. $\sigma(s)$ may be viewed as a Lagrange multiplier which acts locally to constrain the curve derivative at each point (as opposed to a constant Lagrange multipler, which would impose a global constraint).\\

Putting the coefficient of $\delta\mathbf{r}$ in Eqn.~(\ref{eqn:freeEnergyVariations}) to zero gives a differential equation for the shape of the hair
\begin{equation}
K\ddddot{\mathbf{r}}-\frac{d(\sigma\dot{\mathbf{r}})}{ds}=\mathbf{0}\;,   \label{eqn:4ODE} 
\end{equation}
with the boundary conditions
\begin{align}
\mathbf{r}(0)&=\mathbf{0}\;,  \label{eqn:root}  \\
\dot{\mathbf{r}}(0)&=\mathbf{e}_{z}\;,  \label{eqn:clamp}  \\
K\ddot{\mathbf{r}}^{-}(s_{0})&=K\ddot{\mathbf{r}}^{+}(s_{0})\;,   \label{eqn:curvJoin}  \\
\left[K\left(\dddot{\mathbf{r}}^{+}-\dddot{\mathbf{r}}^{-}\right)-\left(\sigma^{+}-\sigma^{-}\right
)\dot{\mathbf{r}}\right]_{s_{0}}&=-\pmb{\gamma}\;,      \label{eqn:forceJoin}  \\
K\ddot{\mathbf{r}}(L)&=\mathbf{0}\;,    \label{eqn:curvEnd}   \\
\left[-K\dddot{\mathbf{r}}+\sigma\dot{\mathbf{r}}\right]_{L}&=\pmb{\gamma}\;. \label{eqn:forceEnd}
\end{align}
Eqns.~(\ref{eqn:root},\ref{eqn:clamp}) are imposed boundary conditions, which arise from fixing the position and orientation of the hair at $s=0$ respectively. Eqns.~(\ref{eqn:curvJoin}-\ref{eqn:forceEnd}) are natural boundary conditions, arising from the boundary terms of Eqn.~(\ref{eqn:freeEnergyVariations}): $\delta\mathbf{r}$, $\delta\dot{\mathbf{r}}$ are free at $s=s_{0},L$ so their coefficients again must be zero.\\

Integrating (\ref{eqn:4ODE}), and using (\ref{eqn:forceJoin}) and (\ref{eqn:forceEnd}) to set the 
constants of integration, gives
\begin{equation}
K\dddot{\mathbf{r}}-\sigma\dot{\mathbf{r}}=\begin{cases}       \label{eqn:3ODE}
\mathbf{0},& 0<s<s_{0}\;, \\
-\pmb{\gamma},
& s_{0}<s<L\;.
\end{cases}
\end{equation}
It is useful to note that the righthand side of the equation is the net force exerted on the hair above the point $s$.  The unknown function $\sigma$ may be eliminated by dotting (\ref{eqn:3ODE}) with $\ddot{\mathbf{r}}$ and using $\ddot{\mathbf{r}}.\dot{\mathbf{r}}=\tfrac{1}{2}d(\dot{\mathbf{r}}^{2})/ds=0$. Integrating a second time, and using Eqns.~(\ref{eqn:curvEnd}) and (\ref{eqn:curvJoin}), gives
\begin{equation}
\frac{1}{2}K\ddot{\mathbf{r}}^{2}(s)=\begin{cases}
\pmb{\gamma.}(\dot{\mathbf{r}}(L)-\dot{\mathbf{r}}(s_{0}))&0<s<s_{0},   \label{eqn:2ODE}  \\
\pmb{\gamma.}(\dot{\mathbf{r}}(L)-\dot{\mathbf{r}}(s))&s_{0}<s<L\;.
\end{cases}
\end{equation}
It is useful at this point to switch to an angular representation. Writing $\dot{\mathbf{r}}=\sin\psi\;\mathbf{e}_{x}+\cos\psi\;\mathbf{e}_{z}$ and $\pmb{\gamma}=\gamma\left(\sin\alpha\;\mathbf{e}_{x}+\cos\alpha\;\mathbf{e}_{z}\right)$, and defining the elastocapillary length $\lambda=\sqrt{\tfrac{K}{\gamma}}$, Eqn.~(\ref{eqn:2ODE}) becomes
\begin{equation}
\tfrac{1}{2}\dot{\psi}^{2}(s)=\begin{cases}
\lambda^{-2}\left(\cos\left[\alpha-\psi(L)\right]-\cos\left[\alpha-\psi(s_{0})\right]\right)&0<s<s_{0}\;,           \\
\lambda^{-2}\left(\cos\left[\alpha-\psi(L)\right]-\cos\left[\alpha-\psi(s)\right]\right)&s_{0}<s<L\;.
\end{cases}
\label{eqn:bendingEqn}
\end{equation}
Eqn.~(\ref{eqn:bendingEqn}) implies that, in the lower segment of the hair, $0<s< s_0$, $\psi(s)$ increases linearly with arc length
\begin{align}
\psi(s)=\lambda^{-1}\sqrt{2\left(\cos\left[\alpha-\psi(L)\right]-\cos\left[\alpha-\psi(s_{0})\right]\right)}s\; .    \label{eqn:linearBending}
\end{align}
For $s_0<s<L$ the nonlinear differential equation for the shape of the hair can be solved in terms of elliptic functions~\cite{ZakharovOkhotkin}. Making the substitution
\begin{equation}
\sin\chi(s)=\frac{\cos\left[\tfrac{1}{2}(\alpha-\psi(s))\right]}{\cos\left[\tfrac{1}{2}(\alpha-\psi(L))\right]}
\end{equation}
in Eqn.~(\ref{eqn:bendingEqn}) gives
\begin{align}
\tfrac{1}{2}\dot{\psi}^{2}&=2\left(\cos^{2}\left[\tfrac{\alpha-\psi(L)}{2}\right]-\cos^{2}\left[\tfrac{\alpha-\psi(s)}{2}\right]\right)\;, \\
2\dot{\chi}^{2}\cos^{2}\chi\frac{\cos^{2}\left[\tfrac{\alpha-\psi(L)}{2}\right]}{\sin^{2}\left[\tfrac{\alpha-\psi(s)}{2}\right]}&=2\lambda^{-2}\cos^{2}\left[\tfrac{\alpha-\psi(L)}{2}\right]\left(1-\sin^{2}\chi\right)\;,\\
\dot{\chi}^{2}&=\lambda^{-2}\left(1-\cos^{2}\left[\tfrac{\alpha-\psi(L)}{2}\right]\sin^{2}\chi\right)\;,
\end{align}
which may be integrated in terms of elliptic integrals
\begin{align}
\lambda^{-1}\left(L-s\right)&=\int_{s}^{L}\frac{d\chi}{\sqrt{1-\cos^{2}\left[\tfrac{\alpha-\psi(L)}{2}\right]\sin^{2}\chi}} \\
&=\mathsf{K}_{\cos\left[\tfrac{1}{2}(\alpha-\psi(L))\right]}-\mathsf{F}_{\cos\left[\tfrac{1}{2}(\alpha-\psi(L))\right]}\left[\chi(s)\right]
\end{align}
where $\mathsf{K}_{m}$ and $\mathsf{F}_{m}$ denote complete and incomplete elliptic integrals of the first kind with modulus $m$~\cite{Wolfram}. The expression inverts as
\begin{equation}
\sin\chi(s)=\frac{\cn_{\cos\left[\tfrac{1}{2}(\alpha-\psi(L))\right]}\left[\lambda^{-1}(L-s)\right]}{\dn_{\cos\left[\tfrac{1}{2}(\alpha-\psi(L))\right]}\left[\lambda^{-1}(L-s)\right]}\; \label{eqn:rodform}
\end{equation}
where $\cn_{m}$ and $\dn_{m}$ are the Jacobi elliptic cosine and delta.\\

$\psi(s)$ must be continuous, so two implicit equations for the constants $\psi(L)$, $\psi(s_0)$ and $\alpha$ follow from taking $s=s_0$ in Eqns.~(\ref{eqn:linearBending}) and~(\ref{eqn:rodform})
\begin{align}
\psi(s_{0})^{2}&=2\left(\frac{s_{0}}{\lambda}\right)^{2}\left(\cos\left[\alpha-\psi(L)\right]-\cos\left[\alpha-\psi(s_{0})\right]\right)\;,  \label{eqn:selfConsistencyBase}  \\ 
\cos\left[\tfrac{1}{2}(\alpha-\psi(s_{0}))\right]&=\cos\left[\tfrac{1}{2}(\alpha-\psi(L))\right]\frac{\cn_{\cos\left[\tfrac{1}{2}(\alpha-\psi(L))\right]}\left[\lambda^{-1}(L-s_{0})\right]}{\dn_{\cos\left[\tfrac{1}{2}(\alpha-\psi(L))\right]}\left[\lambda^{-1}(L-s_{0})\right]}\;. \label{eqn:selfConsistencyJoin}
\end{align}
The third equation determining these parameters follows from the geometry of the model. For the doublet state $\mathbb{P}_{2}$,
symmetry implies $\alpha=\tfrac{\pi}{2}$, provided the hairs do not overlap. For the singlet state
$\mathbb{P}_{1}$ the position of the contact points on successive hairs and the slope of the interface $\alpha$ are related by
\begin{equation}
\cot\alpha=\frac{-\int_{s_{0}}^{L}\cos\psi(s)ds}{D-\int_{s_{0}}^{L}\sin\psi(s)ds}\;, \label{eqn:selfConsistencyAngle}
\end{equation}
which by trignometric rearrangment becomes
\begin{equation}
D\cos\alpha+\int_{s_{0}}^{L}\left(\sin\left[\alpha-\psi(s)\right]\cos^{2}\alpha-\cos\left[\alpha-\psi(s)\right]\cos\alpha\sin\alpha\right)ds=-\int^{L}_{s_{0}}\left(\cos\left[\alpha-\psi(s)\right]\cos\alpha\sin\alpha+\sin\left[\alpha-\psi(s)\right]\sin^{2}\alpha\right)\;.
\end{equation}
Therefore,
\begin{equation}
D\cos\alpha=-\int_{s_{0}}^{L}\sin\left[\alpha-\psi(s)\right]ds\;, \\
\end{equation}
which, by use of the first equality in Eqn.~(\ref{eqn:bendingEqn}), may be written
\begin{equation} 
\begin{split}
D\cos\alpha&=-\int_{\psi(s_{0})}^{\psi(L)}\frac{\lambda\sin\left[\alpha-\psi(s)\right]}{\sqrt{2\left(\cos\left[\alpha-\psi(L)\right]-\cos\left[\alpha-\psi(s)\right]\right)}}d\psi \\
&=-\lambda\sqrt{2\left(\cos\left[\alpha-\psi(L)\right]-\cos\left[\alpha-\psi(s_{0})\right]\right)} \\
&=-\lambda^{2}\frac{\psi(s_{0})}{s_{0}}                    \label{eqn:selfConsistencyAngleNoInt}
\end{split}
\end{equation}
where we have used Eqn.~(\ref{eqn:selfConsistencyBase}) in the final step.\\

Substituting Eqn.~(\ref{eqn:bendingEqn}) into Eqn.~(\ref{eqn:freeEnergyZeroP}), integrating, and
using Eqn.~(\ref{eqn:rodform}), the free energy may now be written.
\begin{equation}
\begin{split}
\frac{\mathcal{F}}{\gamma}=&L\cos\left[\alpha-\psi(L)\right]-s_{0}\cos\left[\alpha-\psi(s_{0})\right]-(L-s_{0})\cos\young+\int_{s_{0}}^{L}\cos\left(\alpha-\psi(s)\right)ds+\lvert\pmb{\Lambda}\rvert-D    \\
=&L\cos\left[\alpha-\psi(L)\right]-s_{0}\cos\left[\alpha-\psi(s_{0})\right]-(L-s_{0})(\cos\young+2)\\
 &\;\;\;\;\;+4\left(\mathsf{E}_{\cos\left[\tfrac{1}{2}(\alpha-\psi(L))\right]}\left(\tfrac{\pi}{2}\right)-\mathsf{E}_{\cos\left[\tfrac{1}{2}(\alpha-\psi(L))\right]}\left(\chi(s_{0})\right)\right)-D\left(1-\sin\alpha\right) \;,          \label{eqn:freeEnergyExplicit}
\end{split}
\end{equation}
where $\mathsf{E}$ is the elliptic integral of the second kind. To find the minimum of the free energy for a given Young angle we now implement the second part of the extremisation process; taking the derivative of $\mathcal{F}$ with respect to $s_{0}$. After some algebra we obtain
\begin{equation}
\frac{d\mathcal{F}}{ds_{0}}=\gamma\left(\cos\left[\alpha-\psi(s_{0})\right]+\cos\young\right)\;,        \label{eqn:equilibrium}
\end{equation}
which is zero when
\begin{equation}
\psi(s_{0})-\alpha+\pi=\young\;.  \label{eqn:young2DHair}
\end{equation}
Hence, as expected, in equilibrium the angle between the interface and the hair is the Young angle. 
Calculating the second derivative of $\mathcal{F}$ leads to the expected condition for a stable equilibrium, that a small advance of the contact line must reduce the contact angle
\begin{equation}
\frac{d\young}{ds_{0}}>0\;.         \label{eqn:stabilityYoung}
\end{equation}
Finally, we have assumed a partially suspended state which will only be stable if the interface remains pinned to the tip of the hair. Invoking the Gibbs' criterion~\cite{Gibbs}, pinning will occur if
\begin{align}
\alpha-\psi(L)<\young\;.      \label{eqn:depin}
\end{align}
Thus the contact angle at the tip may take a finite range of values, as occurs, for example, on the corner of a grooved surface~\cite{Lipowsky}. In conclusion, partially suspended interface configurations, where the interface runs between the tip of a hair and a point part way up the neighbouring hair, are found by substituting Eqn.~(\ref{eqn:young2DHair}) to eliminate $\psi(s_0)$ from Eqns.~(\ref{eqn:selfConsistencyBase}), (\ref{eqn:selfConsistencyJoin}) and (\ref{eqn:selfConsistencyAngleNoInt}) 
\begin{align}
(\alpha+\young-\pi)^{2}&=2\left(\frac{s_{0}}{\lambda}\right)^{2}\left(\cos\left[\alpha-\psi(L)\right]+\cos\young\right)\;,  \label{eqn:selfConsistencyBaseYoung}  \\
\sin\left[\tfrac{1}{2}\young\right]&=\cos\left[\tfrac{1}{2}(\alpha-\psi(L))\right]\frac{\cn_{\cos\left[\tfrac{1}{2}(\alpha-\psi(L))\right]}\left[\lambda^{-1}(L-s_{0})\right]}{\dn_{\cos\left[\tfrac{1}{2}(\alpha-\psi(L))\right]}\left[\lambda^{-1}(L-s_{0})\right]}\;,  \label{eqn:selfConsistencyJoinYoung} \\
\alpha&=\begin{cases}\pi-\young-\lambda^{-2}Ds_{0}\cos\alpha &\;\;\;\text{for }\mathbb{P}_{1}\;, \\
 \tfrac{\pi}{2} &\;\;\;\text{for }\mathbb{P}_{2}\;.
\end{cases}   \label{eqn:selfConsistencyAngleYoung}
\end{align}
These states are viable minima given that the conditions~(\ref{eqn:stabilityYoung}) and (\ref{eqn:depin}) hold.

\section{Singlets}
\label{section:singlets}
\begin{figure}
\centering
\subfigure[]{\label{fig:freeEnergyPlotNeutral}\epsfig{file=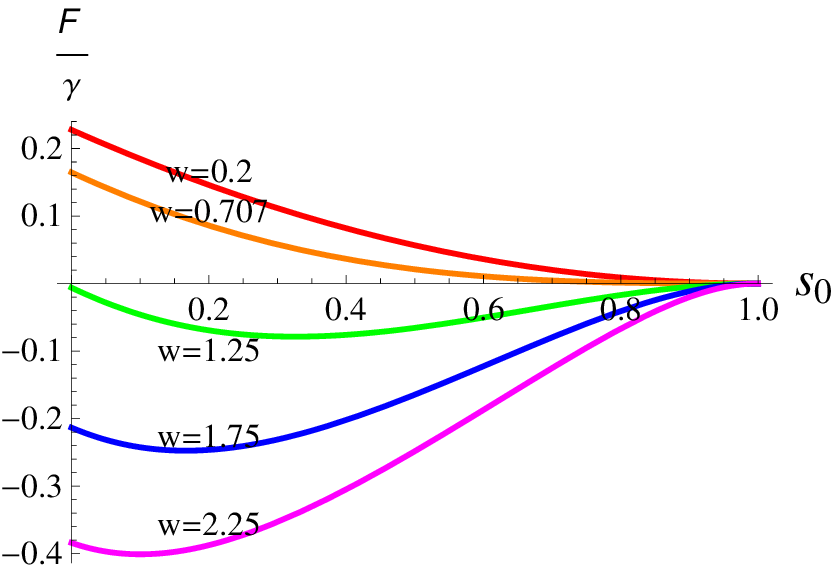,width=55mm}}
\subfigure[]{\label{fig:freeEnergyPlotPhobic}\epsfig{file=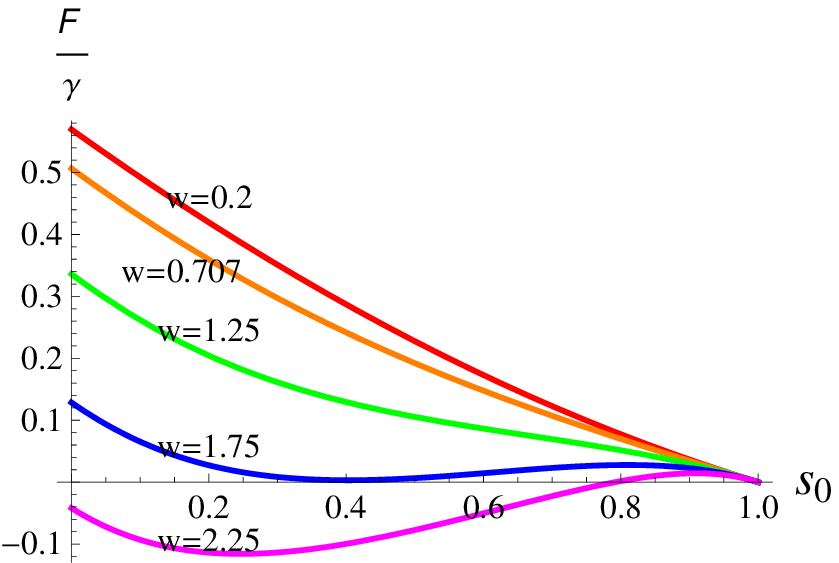,width=55mm}}
\subfigure[]{\label{fig:freeEnergyPlotPhilic}\epsfig{file=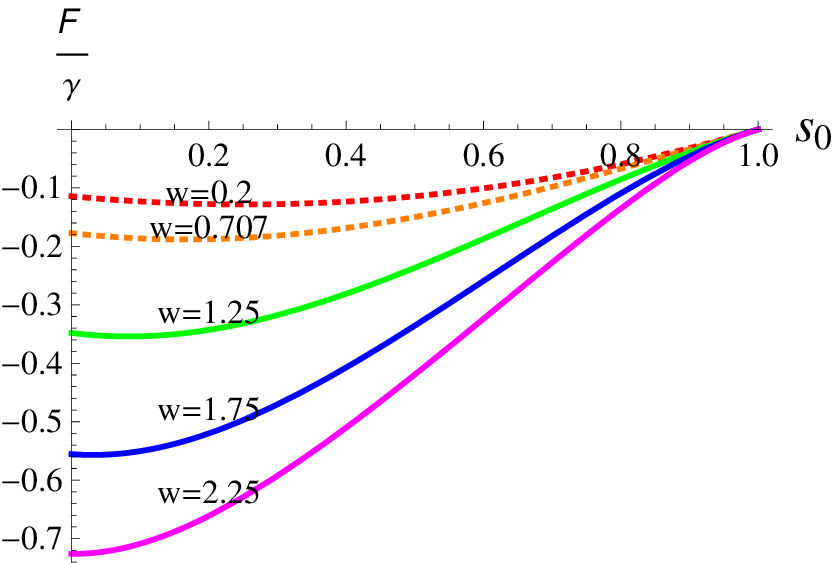,width=55mm}}
\caption{Variation of the reduced free energy $\mathcal{F}/\gamma$ per length $D$, given by Eqn.~(\ref{eqn:freeEnergyExplicit}), with the position 
of the interface-hair contact $s_{0}$ for the partially suspended singlet state $\mathbb{P}_{1}$.
Results for $w=\sqrt{\gamma /K}L$ $=0.25$ (red), $1/\sqrt{2}$ (orange), $1.25$ (green), $1.75$ (blue) and $2.25$
(magenta) are shown for different values of the equilibrium contact angle (a) $\young=90^{\circ}$, (b) $\young=110^{\circ}$, (c) $\young=70^{\circ}$}.
\label{fig:freeEnergySinglets}
\end{figure}
\begin{figure}
\centering
\subfigure[]{\label{fig:phaseDiagramSinglets}\epsfig{file=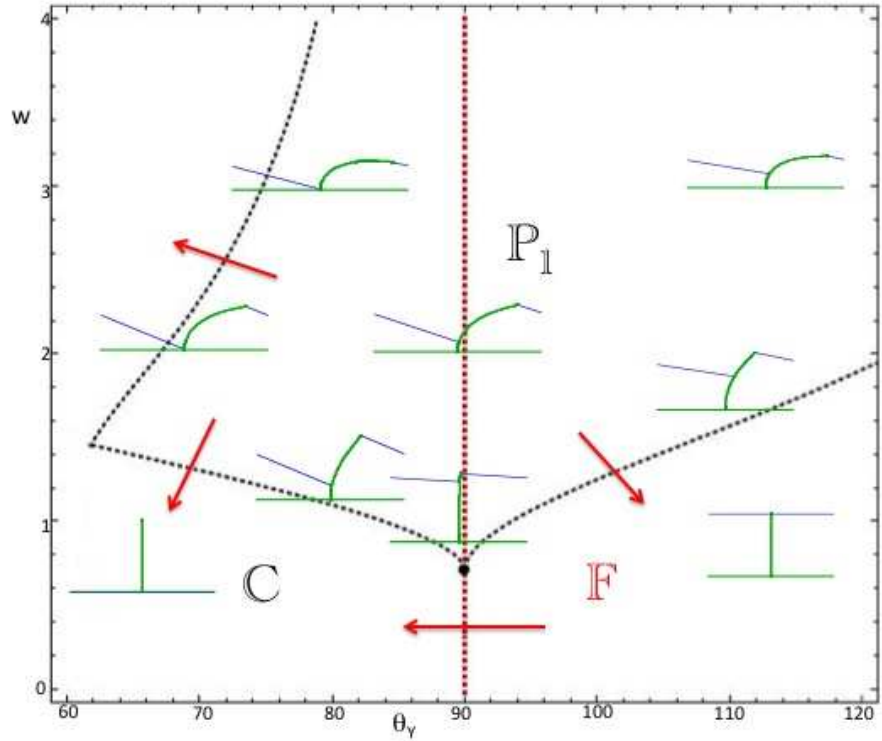,width=120mm}}
\subfigure[]{\label{fig:phaseDiagramDoublets}\epsfig{file=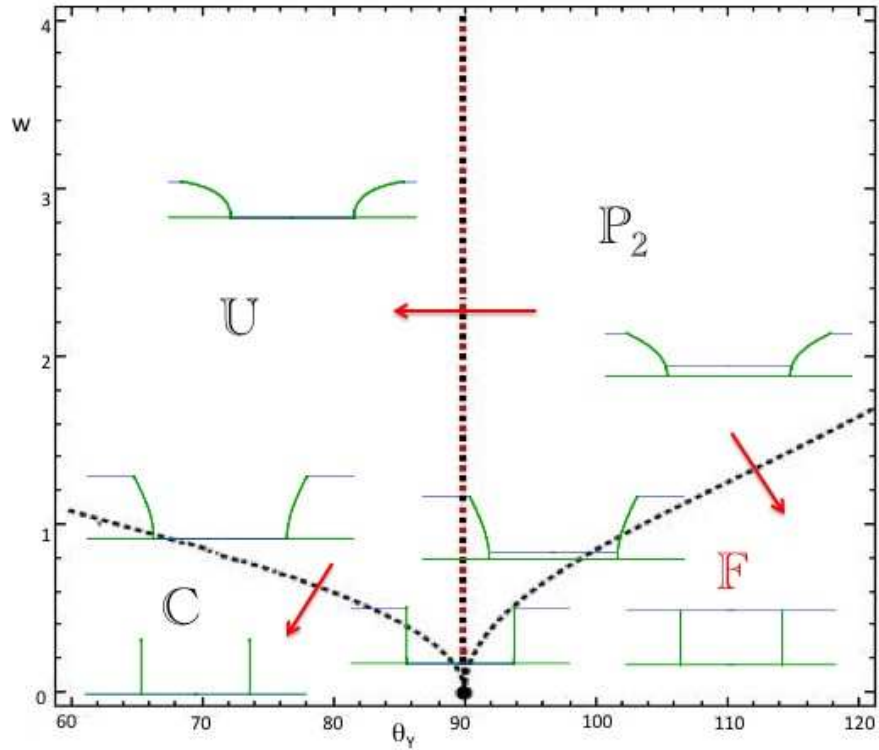,width=120mm}}
\caption{Phase space diagrams for $D/L=2$.  (a) Black dashed lines show the limits of 
(meta)stability of the partially suspended singlet state $\mathbb{P}_{1}$ and red dashed lines the 
limit of stability of the fully suspended state $\mathbb{F}$.  Note that $\mathbb{P}_{1}$ can 
persist at hydrophilic contact angles. (b) Black dashed lines show the limits of (meta)stability of 
the partially suspended doublet state $\mathbb{P}_{2}$ and of the unzipped state  $\mathbb{U}$, and
red dashed lines the limit of stability of $\mathbb{F}$. Boundaries correspond to transitions which 
take place in the direction of the arrows.
 $w=\sqrt{\gamma/K}L$ is a dimensionless parameter describing how easily the hairs can be bent by 
the fluid interface and $\young$ is the equilibrium contact angle of the hairs. Diagrams have been 
added which show the hair and interface profiles at given positions in parameter space: hairs are 
depicted in green (lighter grey) and the liquid interface in blue (darker grey).}
\label{fig:phaseDiagrams}
\end{figure}
We first assume that the drop is in the partially suspended singlet state $\mathbb{P}_{1}$ and discuss when this state can remain metastable with respect to the fully suspended and collapsed configurations $\mathbb{F}$ and $\mathbb{C}$. A similar analysis for the partially suspended doublet is given in the next section. The results are summarised in Fig.~\ref{fig:phaseDiagramSinglets} and are presented in terms of the dimensionless parameter $w:=\lambda^{-1}L$
which is a measure of the ability of the interface to bend the hairs; $w=0$ for rigid hairs.\\

Consider first neutral wetting, $\young=90^{\circ}$. The variation of the free energy of the partially suspended state $\mathbb{P}_{1}$ is shown in Fig.~\ref{fig:freeEnergyPlotNeutral} for different values of $w$. Recalling that the suspended state $\mathbb{F}$ corresponds to $s_{0}=1$, and that its free energy is chosen as zero, the curves show that, for small $w$, $\mathbb{F}$ is stable. For $w>w_{0}$, however, $s_{0}=1$ becomes unstable, and a new stable minimum, corresponding to $\mathbb{P}_1$ appears. Its position decreases continuously from $s_0=1$ as $w$ increases from $w_0$ signalling a continuous transition. In the Appendix we show that $w_0=\sqrt{L/D}$.\\

There is similar behaviour in the hydrophobic region of parameter space, as shown by the free energy curves in Fig.~\ref{fig:freeEnergyPlotPhobic} for $\young=110^\circ$. A drop in the partially suspended singlet state will rise to be fully suspended for small $w$, but will remain in $\mathbb{P}_1$ for larger $w$. The new feature for $\young>90^\circ$ is that the free energy has a maximum near $s_0=1$ indicating that the $\mathbb{P}_1 \rightarrow \mathbb{F}$ transition is not reversible. The boundary of metastability moves to higher $w$ with increasing $\young$ as shown in Fig.~\ref{fig:phaseDiagramSinglets} for $D/L=2$.\\

A typical free energy plot for hydrophilic hairs in the partially suspended state is shown in Fig.~\ref{fig:freeEnergyPlotPhilic}. For hydrophilic hairs, $\mathbb{P}_{1}$ is susceptible to collapse by two distinct mechanisms, which place upper and lower limits of stability on the state.For large $w$ there is no stable $\mathbb{P}_1$ state.The interface will descend to $s_0=0$ and then spread across the substrate to form a collapsed state.  For smaller $w$ the free energy curves have a minimum corresponding to a $\mathbb{P}_{1}$ state that is stable to collapse. The minimum moves continuously to $s_0=0$ as $w$ increases but the transition is not reversible as there is a strong free energy barrier associated with dewetting the base substrate. Alternatively, if $w$ is decreased to $0$, the minimum in the free energy (\ref{eqn:freeEnergyExplicit}) persists which is, at first sight, odd because $\mathbb{P}_1$ is not stable in the rigid limit. This is resolved by noting that the Gibbs' criterion~(\ref{eqn:depin}) is violated for small $w$. The partially suspended singlets are no longer stable and there is a first order transition to the collapsed phase. Free energy curves which correspond to these values of $w$ are indicated by dotted lines in Fig.~\ref{fig:freeEnergyPlotPhilic}. The explicit form of the depinning curve may be found by substituting Eqn.~(\ref{eqn:depin})  (written as an equality) into Eqns.~(\ref{eqn:selfConsistencyBaseYoung}-\ref{eqn:selfConsistencyAngleYoung}), and eliminating $\alpha$ and $s_{0}$, 
\begin{equation}
\frac{L}{\lambda}=\frac{1}{2\sqrt{\cos\young}}\left(\arccos\left[-2\frac{\lambda}{D}\sqrt{\cos\young}\right]+\young-\pi\right)+\mathsf{K}_{\cos\tfrac{\young}{2}}-\mathsf{F}_{\cos\tfrac{\young}{2}}\left(\arcsin[\tan\tfrac{\young}{2}]\right)\;,  \label{eqn:depinExplicit}
\end{equation}
where $\mathsf{K}_{m}$ and $\mathsf{F}_{m}$ denote complete and incomplete elliptic integrals of the first kind with modulus $m$. The boundary where $\mathbb{P}_1$ becomes unstable to $\mathbb{C}$ changes direction at a prominent cusp marking the crossover between the two different collapse mechanisms. The cusp is significant because it marks the lowest $\young$ for which $\mathbb{P}_{1}$ is metastable for a given $L/D$. The position of the cusp can be found by substituting the depinning condition~(\ref{eqn:depin}) and $s_{0}=0$ into  Eqns.~(\ref{eqn:selfConsistencyBaseYoung}-\ref{eqn:selfConsistencyAngleYoung}), giving the equalities
\begin{equation}
\frac{L}{\lambda}=\mathsf{K}_{\cos\tfrac{\young}{2}}-\mathsf{F}_{\cos\tfrac{\young}{2}}\left(\arcsin\left[\tan\tfrac{\young}{2}\right]\right)=\frac{2 L}{D\sqrt{\cos\young}} .
\label{eqn:cusp}
\end{equation}\\
We now describe how the metastability of the partially suspended singlet state varies with the hair spacing to length 
ratio $D/L$.
$w_{0}$, the position of the continuous $\mathbb{P}_{1}\leftrightarrow\mathbb{F}$ transition for neutral wetting, decreases as $D$ increases thus making $\mathbb{P}_{1}$ states accessible to hairs of greater rigidity or,
equivalently, to liquids of lower surface tension. In contrast, the range of hydrophilic Young
angles for which $\mathbb{P}_{1}$ is metastable is reduced, with the cusp in the
$\mathbb{P}_{1}\rightarrow\mathbb{C}$ curve moving to higher contact angles. Indeed, in the limit $D\to\infty$, Eqns.~(\ref{eqn:criticalPoint}) and (\ref{eqn:cusp}) show that the cusp and critical point converge at ($\young=90^{\circ},w=0$), and there are no partially suspended states in the hydrophilic region. \\

We consider another possible drop configuration, $\mathbb{Q}_1$, where both ends of a segment of interface lie part way down the posts. This state could, for example, be created when an interface in the $\mathbb{P}_1$ configuration depins from the tip of the hairs. In addition to the contact point at $s=s_{0}$, the second contact point in a $\mathbb{Q}_{1}$ state will be situated at a higher point $s=\tilde{L}$. Above the two contact points the hair will experience no forces and hence will not be bent. Therefore equilibrium conditions for $\mathbb{Q}_{1}$ can be obtained by making the substitutions $L\rightarrow\tilde{L}$ in Eqns.(\ref{eqn:selfConsistencyBaseYoung}-\ref{eqn:selfConsistencyAngleYoung}). However, the analogy is not complete because Eqn.~(\ref{eqn:depin}), which represented the Gibbs' criterion for $\mathbb{P}$ states, is now an equality. Thus $\mathbb{Q}_{1}$ states obey Eqn.~(\ref{eqn:depinExplicit}). In order for these equilibria to be stable we require $\tfrac{d\theta_{Y}}{d\tilde{L}}>0$, in analogy with Eqn.~(\ref{eqn:stabilityYoung}), or equivalently that $\tfrac{d\tilde{L}}{d\theta_{Y}}>0$. However, examination of the righthand side of Eqn.~(\ref{eqn:depinExplicit}) shows it to be a monotonically decreasing function of $\young$ for all values of $\lambda/D$, and therefore there are no stable $\mathbb{Q}_{1}$ states.\\

Finally, having found stable equilibria for a systems in which periodicity is enforced, we should check that such states are not unstable to variations between hair profiles. To do this we consider a more general system, where each hair $n$ has an independent profile, and check the stability of the uniform states, given by Eqs.~(\ref{eqn:selfConsistencyBaseYoung}-\ref{eqn:selfConsistencyAngleYoung}), against perturbations of all possible wavelengths along the chain of hairs. We find stability for all $\mathbb{P}_{1}$ states, except for a small region along the $\mathbb{F}\rightarrow\mathbb{P}_{1}$ curve close to the critical point. For the system considered here, this region extends only $\sim 1^{\circ}$.

\section{Doublets}
\label{section:doublets}

We now analyse the stability of the partially suspended doublet states $\mathbb{P}_2$ against transitions to $\mathbb{F}$, $\mathbb{C}$ and the unzipped state $\mathbb{U}$ (see Fig.~\ref{fig:2Dstates}). These are easier to handle because the additional symmetry implies that the interface lies parallel to the surface, so $\alpha=\tfrac{\pi}{2}$ and the results are independent of $D$ (provided that the hairs are sufficiently spaced not to touch).\\

The equations of equilibrium follow from substituting $\alpha=\tfrac{\pi}{2}$ into Eqns.~(\ref{eqn:selfConsistencyBaseYoung}) and 
(\ref{eqn:selfConsistencyJoinYoung}), giving\\
\begin{align}
\left(\young-\frac{\pi}{2}\right)^{2}&=2\left(\frac{s_{0}}{\lambda}\right)^{2}\left(\sin\psi(L)+\cos\young\right)\;,  \label{eqn:selfConsistencyDoubletBase}  \\
\sin\left[\tfrac{1}{2}\young\right]&=\cos\left[\tfrac{1}{2}\left(\tfrac{\pi}{2}-\psi(L)\right)\right]\frac{\cn_{\cos\left[\tfrac{1}{2}\left(\tfrac{\pi}{2}-\psi(L)\right)\right]}\left[\lambda^{-1}\left(L-s_{0}\right)\right]}{\dn_{\cos\left[\tfrac{1}{2}\left(\tfrac{\pi}{2}-\psi(L)\right)\right]}\left[\lambda^{-1}\left(L-s_{0}\right)\right]}\;. 
\label{eqn:selfConsistencyDoubletJoin}
\end{align}
A phase diagram for doublets, constructed by solving Eqns.~(\ref{eqn:selfConsistencyDoubletBase}) and
(\ref{eqn:selfConsistencyDoubletJoin}), together with the stability condition~(\ref{eqn:stabilityYoung})
and the depinning inequality~(\ref{eqn:depin}) is shown in Fig.~\ref{fig:phaseDiagramDoublets}. \\

The hydrophobic portion of the phase diagram is similar to that for singlets: $\mathbb{P}_{2}$ undergoes an irreversible transition to $\mathbb{F}$ as $w$ decreases. The region over which $\mathbb{P}_{2}$ is stable increases with decreasing contact angle. The partially suspended doublet exists only for hydrophobic hairs because $s_{0}\rightarrow 0$ as $\young \rightarrow 90^{\circ}$. At the $\young=90^{\circ}$ boundary  $\mathbb{P}_{2}$ becomes unstable to the unzipped state $\mathbb{U}$. This in turn collapses to $\mathbb{C}$ for smaller $\young$, when the bridging sections of interface in $\mathbb{U}$ depin from the tips of the hairs.\\

\section{Producing the partially suspended states}
\label{section:latticeBoltzmann}
\begin{figure}
\centering
\subfigure[]{\label{fig:uprightLB}\epsfig{file=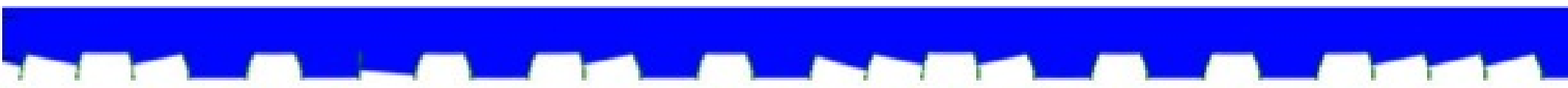,width=180mm}}
\subfigure[]{\label{fig:inclinedLB}\epsfig{file=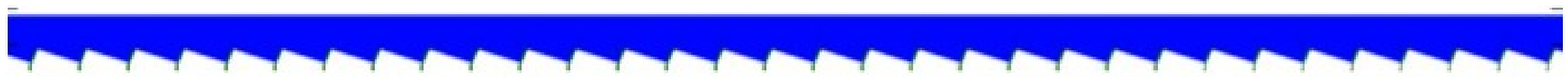,width=180mm}}
\caption{Lattice Boltzmann simulations showing the final configuration of an interface which is 
gently placed in the suspended state $\mathbb{F}$ on (a) vertical hairs, (b) hairs inclined at an 
angle of $5^{\circ}$ to the vertical. Even a small angle of inclination favours the formation of a 
uniform, partially suspended singlet state. Parameters are $w=0.98$, $\young=90^{\circ}$.}
\label{fig:LBsimulations}
\end{figure}

In Secs.~\ref{section:singlets} and~\ref{section:doublets}, we discussed the stabilty of the partially suspended states {\em assuming that the system was initially in these configurations}. We found that these states can remain stable against transitions to $\mathbb{F}$, $\mathbb{C}$ or $\mathbb{U}$ over a substantial range of parameter space. In particular the partially suspended singlet persists at hydrophilic contact angles. We now address the question of how easily the partially suspended states can be created in the first place. \\

Consider a large drop in the fully suspended state, $\mathbb{F}$. If the Young angle is not hydrophobic, $\mathbb{F}$ is unstable and the system must undergo one of three available discontinuous transitions: to $\mathbb{P}_{1}$, $\mathbb{U}$ or $\mathbb{C}$. We find that, for vertical hairs, the chosen state depends very sensitively on any initial perturbation. We demonstrate this using simulations, where the fluid and hairs are modelled using the free energy lattice Boltzmann method and the lattice spring method respectively. Details of these techniques are given in~\cite{Swift,Briant,Buxton}. In Fig.~\ref{fig:uprightLB} the interface is initially placed in $\mathbb{F}$ for $\young=90^{\circ}$ and allowed to evolve with time. A random sequence of regions of $\mathbb{P}_{1}$ and $\mathbb{U}$ result as a consequence of machine noise. \\

This suggests that to robustly produce the partially suspended singlet configuration it would be helpful to impose a geometry which favours the formation of this state. This can be done by tilting the base of the hairs with respect to the vertical.\\
 
In Fig.~\ref{fig:inclinedLB} we show results obtained by repeating the simulation in Fig.~\ref{fig:uprightLB}, but for hairs initially tilted at $5^{\circ}$. The effect of even this slight inclination is to form a uniform partially suspended singlet configuration. We therefore now extend the analysis presented in Secs.~\ref{section:singlets} and~\ref{section:doublets} to show how the region of stability of the $\mathbb{P}_{1}$ configuration is changed by aligning the base of the hairs at an angle $\Omega$ to the vertical.\\

\section{Inclined hairs}
\label{section:2Dincline}
\begin{figure}
\centering
{\epsfig{file=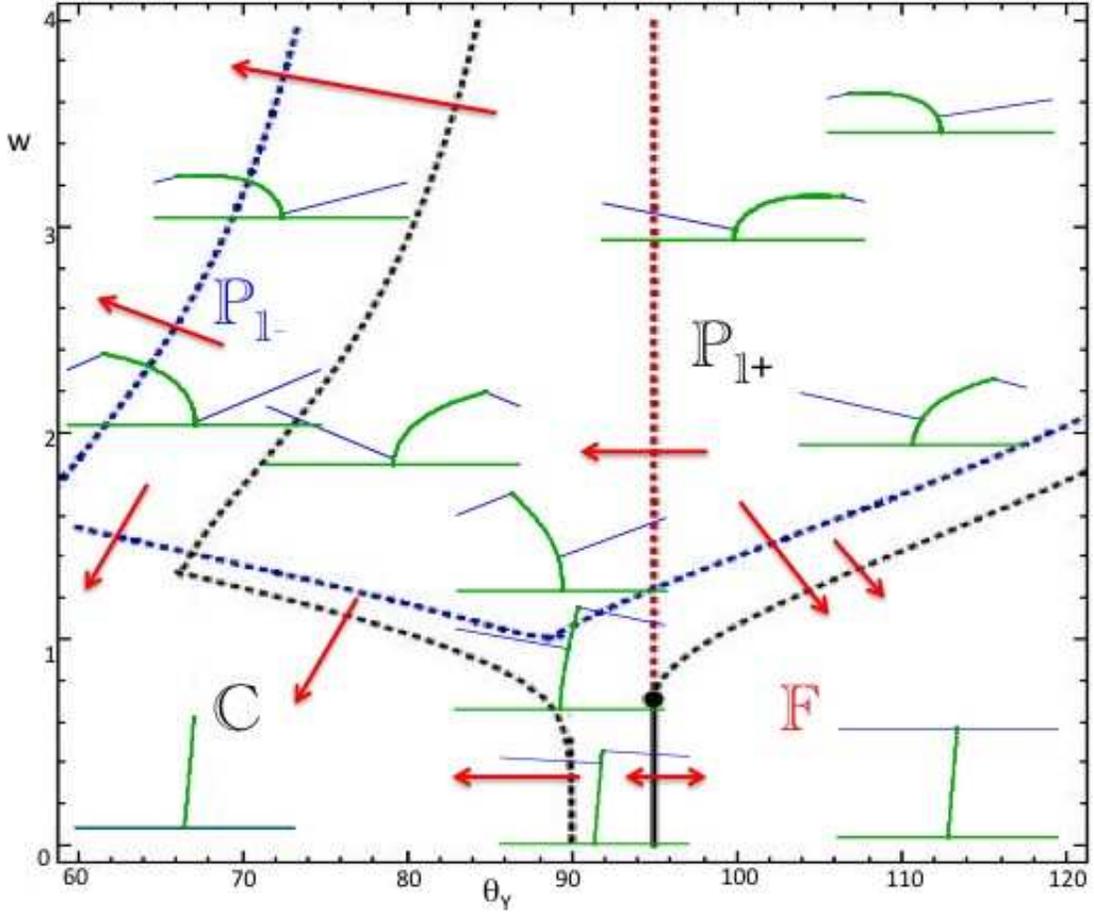,width=150mm}}
\caption{Phase space diagrams for $D/L=2$ and a hair inclination of $5^{\circ}$. Black and blue 
lines show the limits of (meta)stability of the partially suspended singlet states 
$\mathbb{P}_{1+}$ and $\mathbb{P}_{1-}$ respectively, and red lines the limit of stability of the 
fully suspended state $\mathbb{F}$.  
Dotted (solid) lines show the boundaries for discontinuous (continuous) transitions, which take place in the directions of the arrows. $w=\sqrt{\gamma/K}L$ is 
a dimensionless parameter describing how easily the hairs can be bent by the fluid interface and 
$\young$ is the equilibrium contact angle of the hairs. Diagrams have been added which show the 
hair and interface profiles at given positions in parameter space: the hairs bending to the right 
are in the  $\mathbb{P}_{1+}$ state and those to the left in $\mathbb{P}_{1-}$.}
\label{fig:phaseDiagramSingletsInclined}
\end{figure}

We consider hairs which have a fixed inclination $\Omega$ (in the positive $x$ direction, say) from the vertical at their base. Note that the reflection symmetry that holds for $\Omega=0$ is now broken, and we must distinguish two sets of partially suspended singlets, $\mathbb{P}_{1+}$ and $\mathbb{P}_{1-}$, in which the interface between hairs $n$ and $n+1$ is  pinned at the top of hairs $n$ or $n+1$ respectively.\\

For singlets a non-zero inclination angle can be accounted for by a simple modification of Eqns.~(\ref{eqn:selfConsistencyBaseYoung}) and (\ref{eqn:selfConsistencyAngleYoung})
\begin{align}
(\alpha+\young-\pi\mp\Omega)^{2}&=2\lambda^{-2}s_{0}^{2}\left(\cos\left[\psi(L)-\alpha\right]+\cos\young\right)\;,  \label{eqn:selfConsistencyBaseYoungInc}  \\ 
\alpha+\young-\pi\mp\Omega&=-\lambda^{-2}Ds_{0}\cos\alpha\;, \label{eqn:selfConsistencyAngleYoungInc} 
\end{align}
with (\ref{eqn:selfConsistencyJoinYoung}) unchanged. The functional form of the free energy (\ref{eqn:freeEnergyExplicit}) remains the same.\\

Consider first rigid hairs, $w=0$. Simple arguments based on the Gibbs' pinning criterion show that $\mathbb{F}$ is
stable for $\young>90^\circ+\Omega$, while collapse occurs for $\young<90^\circ$. Between these values the free energy is locally minimised by $\mathbb{P}_{1+}$ with the unpinned end of the interface lying at
\begin{align}
s_{0}=L - D\frac{\cos\left[\Omega-\young\right]}{\sin\young}\;.    \label{eqn:rigidInclined}
\end{align}
$\mathbb{P}_{1-}$ states do not exist in the rigid limit. \\

The stability of the partially suspended singlet configurations for flexible hairs is summarised by 
Fig.~\ref{fig:phaseDiagramSingletsInclined} for $D/L=2$ and $\Omega=5^\circ$. Comparing 
Fig.~\ref{fig:phaseDiagramSinglets} for vertical hairs, an important difference is the stability of 
$\mathbb{P}_{1+}$ at $w=0$. For $w<w_{0}$ the transition between $\mathbb{F}$ and $\mathbb{P}_{1+}$ 
is continuous, and hence reversible, and occurs along the line $\young=90^\circ+\Omega$. For 
$w>w_{0}$ the boundaries of metastability of the two phases are different: $\mathbb{F}$ becomes 
unstable to $\mathbb{P}_{1+}$ along $\young=90^\circ+\Omega$ whereas $\mathbb{P}_{1+}$ remains 
stable against a transition to $\mathbb{F}$ for higher $\young$ as $w$ increases.\\

The hydrophilic region of the phase diagram is qualitatively similar for vertical and inclined hairs. However note that the $\mathbb{P}_{1+}$ states are stable less far into the hydrophillic region than for the upright hairs. The $\mathbb{P}_{1-}$ configurations are, unsurprisingly, nearly always of higher free energy than the $\mathbb{P}_{1+}$. However they do extend, as metastable states, further into the hydrophilic region.\\

\section{The effect of Laplace Pressure}
\label{section:LaplacePressure}
\begin{figure}
\centering
{\epsfig{file=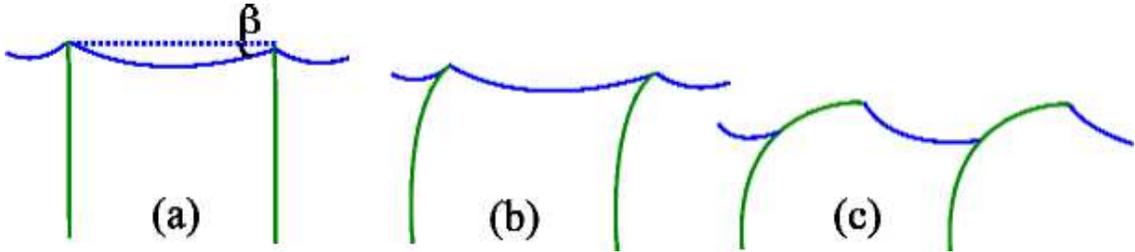,width=150mm}}
\caption{Schematic illustration of the buckling tranisition when a Laplace pressure is applied across the interface, for an increasing $w$: (a) for $w<w_{\mathrm{B}}$ the vertical hairs do not buckle. (b) for $w>w_{\mathrm{B}}$ the hairs buckle but the contact line may remain pinned in an $\mathbb{F}$ state. (c) for a higher $w$, the contact line depins from the hair tip, leading e.g. to a $\mathbb{P}$ state. }
\label{fig:laplace}
\end{figure}

Having determined the equilibrium morphologies for our system, it is interesting to consider how a pressure difference
between the two phases affects the system. An excess pressure in the liquid phase could represent the hydrostatic pressure
experienced by a diving insect, or result from a finite drop size: Laplace's law relates the curvature of the interface $R^{-1}$ to the pressure difference across it, through $p=\gamma R^{-1}$.\\

Departures of the interface from planar, together with the pressure directly applied to the wetted part of the hair, complicate the equations describing the hair profile. However two basic effects of Laplace pressure on the phase diagrams may be understood. Firstly, the transition line for $\mathbb{F}\rightarrow\mathbb{C}$, moves into the hydrophobic region~\cite{Kusumaatmaja}. This is because the bulging interface meets the hair at an angle $90^{\circ}+\beta$ (see Fig.~\ref{fig:laplace} (a)) where
\begin{equation}
\sin\beta=\tfrac{1}{2}DR^{-1}\;,    \label{eqn:beta}
\end{equation}
so Gibbs' pinning is overcome for  $\young<90^{\circ}+\beta$. For elastic hairs, the critical point is also shifted, to $90^{\circ}+\beta+\Omega$, and the value of $w_{0}$ is slightly changed.\\

A second consequence is that $\mathbb{F}$ states will undergo buckling. This is because there is now a downward force $pD$ acting on the tips of the hairs. The profile of the hairs in the fully suspended state obeys a differential equation analogous to (\ref{eqn:bendingEqn})
\begin{equation}
\tfrac{1}{2}K\dot{\psi}^{2}(s)=pD\left(\cos[\pi-\psi(L)]-\cos[\pi-\psi(s)]\right)\;,
\end{equation}
which can be solved to give a condition analogous to (\ref{eqn:selfConsistencyJoin})
\begin{equation}
\sin\left[\frac{\Omega}{2}\right]=\sin\left[\frac{\psi(L)}{2}\right]\frac{\cn_{\sin\left[\frac{\psi(L)}{2}\right]}\left[w\sqrt{2\sin\beta}\right]}{\dn_{\sin\left[\frac{\psi(L)}{2}\right]}\left[w\sqrt{2\sin\beta}\right]}\;.\label{eqn:fullySuspendedLaplace}
\end{equation}
When $\Omega=0$, Eqn.~(\ref{eqn:fullySuspendedLaplace}) implies that either $\psi(L)=0$ or $\cn_{\sin\left[\frac{\psi(L)}{2}\right]}\left[w\sqrt{2\sin\beta}\right]=0$. The second condition is equivalent to
\begin{equation}
w\sqrt{2\sin\beta}=\mathsf{K}_{\sin\left[\frac{\psi(L)}{2}\right]}\;,
\end{equation}
which can be achieved when 
\begin{equation}
w>w_{\mathrm{B}}=\frac{\pi}{\sqrt{8\sin\beta}}\;.
\end{equation}
This is Euler's buckling formula. Above $w_{\mathrm{B}}$, the $\psi(L)=0$ solution to Eqn.~(\ref{eqn:fullySuspendedLaplace}) is unstable and the hairs in $\mathbb{F}$ will be bent. For the case $\Omega\neq 0$, Eqn.~(\ref{eqn:fullySuspendedLaplace}) indicates that there will be a finite degree of bending for any finite $w$. Should the hairs bend sufficiently, such that $\psi(L)+\tfrac{\pi}{2}+\beta>\young$, the interface will depin from the tip, leading to a $\mathbb{F}\rightarrow\mathbb{P}$ transition.

\section{A three dimensional model}
\label{section:3Dmodel}
\begin{figure}
\centering
{\epsfig{file=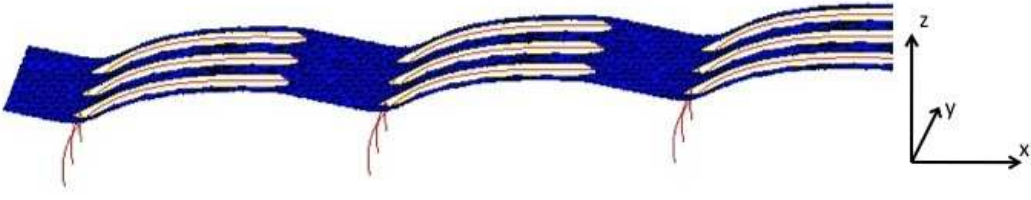,width=140mm}}
\caption{Surface evolver simulations showing the interface and hair profiles for a rectangular 
array of hairs with spacing $D_{x}=2L$, $D_{y}=4a$ and
$L=32a$. $w=3$ and $\young=90^{\circ}$. The interface closely conforms to the bending profile of the hair, with little curvature along $y$.}
\label{fig:surfaceEvolverHairs}
\end{figure}
\begin{figure}
\centering
\subfigure[]{\label{fig:phaseDiagramSinglets3D}\epsfig{file=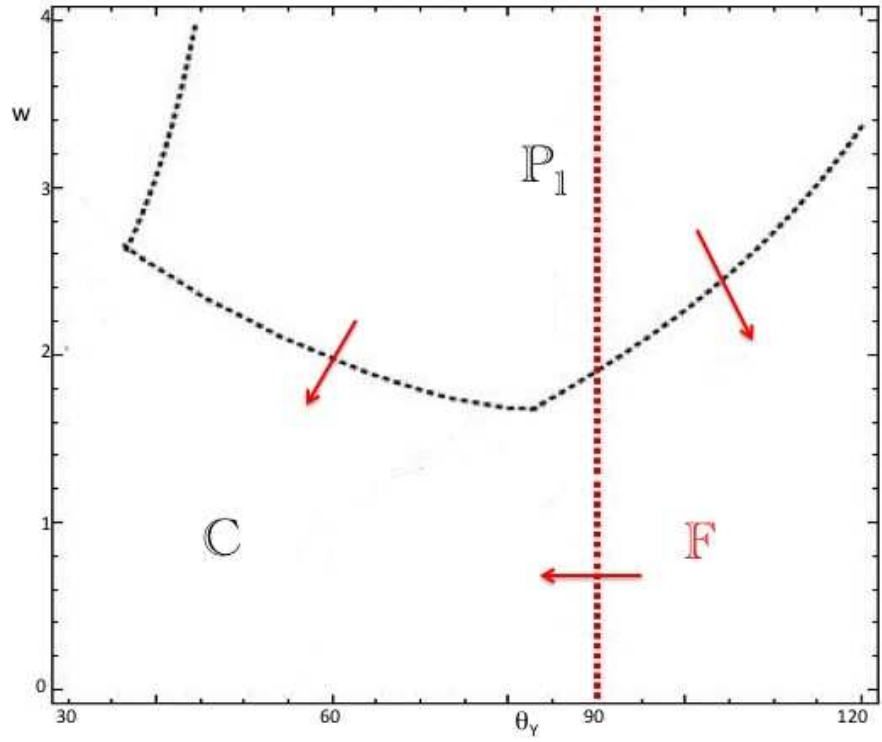,width=120mm}}
\subfigure[]{\label{fig:phaseDiagramDoublets3D}\epsfig{file=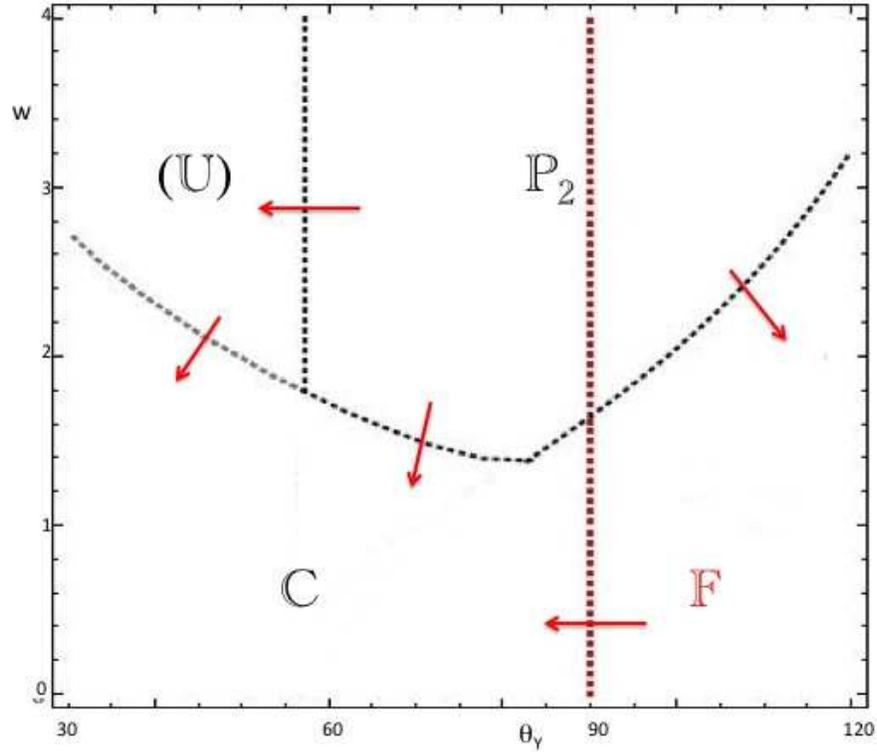,width=120mm}}
\caption{Phase space diagrams for hair spacing $D_{x}=2L$, $D_{y}=4a$ and radius $a=L/32$. (a) 
Black dashed lines show the limits of (meta)stability of the partially suspended singlet state 
$\mathbb{P}_{1}$ and red dashed lines the limit of stability of the fully suspended state 
$\mathbb{F}$.  (b) Black dashed lines show the limits of (meta)stability of the partially suspended 
doublet state $\mathbb{P}_{2}$ and of the unzipped state  $\mathbb{U}$, and
red dashed lines the limit of stability of $\mathbb{F}$. Boundaries correspond to transitions which 
take place in the direction of the arrows.
$w=\sqrt{\gamma D_{y}/K}L$ is a dimensionless parameter describing how easily the hairs can be bent 
by the fluid interface and $\young$ is the equilibrium contact angle of the hairs. Comparing with 
Fig.~\ref{fig:phaseDiagrams}, we see that $\mathbb{P}_{1}$ and $\mathbb{P}_{2}$ can exist at much 
lower contact angles than for the two dimensional geometry, but only at higher $w$.}
\label{fig:phaseDiagrams3D}
\end{figure}
We now consider an infinite drop lying on a two-dimensional array of hairs. The hairs are anchored to the substrate at $z=0$ to form a rectangular lattice, with separation $D_{x}$ and $D_{y}$ in the $x$ and $y$ directions. It is no longer meaningful to model the hairs as having zero width, and they are ascribed a constant radius $a$. This three dimensional geometry includes additional physics that results from a connected, two-dimensional interface. However, to make analytic progress building on our previous results, we need to keep the artificial constraint that the hairs can bend only in the $xz$ plane so that their spacing along $y$ remains constant.\\

Partially wetting cylinders are bound to an interface because it is energetically favourable for them to lie there rather than in either bulk phase. If a cylinder is placed on a flat interface then it will adjust its position relative to the interface to achieve the correct equilibrium contact angle $\young$. Similarly elastic hairs can be bound to an interface, as the decrease in wetting and interfacial contributions to the free energy can be greater than the penalty due to bending.\\

We restrict ourselves to geometries where the interface curvature along $y$ can be neglected. In general, we expect this to be a good approximation for $(\lambda,L,D_{x})$ large compared to $D_{y}$ as borne out by the simulated hair profiles shown in Fig.~\ref{fig:surfaceEvolverHairs}. These results, obtained using Surface Evolver~\cite{Brakke}, show that for $D_{x}=2L$, $D_{y}=4a$ and hair radius $a=L/32$, the interface conforms closely to the bending profile of the hair with little variation in the $y$-direction.\\

Under this approximation, we can write down a free energy analogous to Eqn.~(\ref{eqn:freeEnergyZeroP})
\label{section:2Dzero}
\begin{multline}
\mathcal{F}=\int_{0}^{s_{0}}\left[\tfrac{1}{2}K\ddot{\mathbf{r}}^{2}+\tfrac{1}{2}\sigma(s)\left(\dot{\mathbf{r}}^{2}-1\right)\right]ds+\int_{s_{0}}^{L}\left[\tfrac{1}{2}K\ddot{\mathbf{r}}^{2}+\tfrac{1}{2}\sigma(s)\left(\dot{\mathbf{r}}^{2}-1\right)-\gamma\cos\young2a(\pi-\young)\right]ds\\
+\gamma\left[D_{y}(\vert\pmb{\Lambda}\vert-D_{x})+\int^{L}_{s_{0}}(D_{y}-2a\sin\young)ds\right].   \label{eqn:freeEnergyZeroP3D}
\end{multline}
The parts of the free energy (\ref{eqn:freeEnergyZeroP3D}) relating to bending are unchanged in their functional form. The wetting contribution, in the second integrand, takes account of the hair making an angle $\young$ with the interface, such that an area $2a(\pi-\young)$ per unit length of the hair is wetted. The interfacial part of the free energy, the last term of (\ref{eqn:freeEnergyZeroP3D}), contains two contributions. The first is from the area of the interface between the tips of one row of hairs, and the contact point $s_{0}$ of the next row in the $x$ direction, assuming negligible variation of the interface in the $y$ direction. We recognise that this part is the same as the interfacial term in Eqn.~(\ref{eqn:freeEnergyZeroP}), but with $\gamma$ replaced by $D_{y}\gamma$, reflecting the change of dimension. The second contribution arises from the interface in the gaps between neighbouring hairs in the $y$ direction, again taking into account that the interface meets the hair at the angle $\young$. We note that this term may be assimilated into the wetting term, so that the free energy in Eqn. (\ref{eqn:freeEnergyZeroP3D}) may be rewritten as
\begin{equation}
\mathcal{F}=\int_{0}^{s_{0}}\left[\tfrac{1}{2}K\ddot{\mathbf{r}}^{2}+\tfrac{1}{2}\sigma(s)\left(\dot{\mathbf{r}}^{2}-1\right)\right]ds+\int_{s_{0}}^{L}\left[\tfrac{1}{2}K\ddot{\mathbf{r}}^{2}+\tfrac{1}{2}\sigma(s)\left(\dot{\mathbf{r}}^{2}-1\right)-\gamma D_{y}\cos\cassie\right]ds+\gamma D_{y}(\vert\pmb{\Lambda}\vert-D_{x})\;,   \label{eqn:freeEnergyZeroP3Db}
\end{equation}
where
\begin{equation}
\cos\cassie=-1+2\frac{a}{D_{y}}\left[\sin\young+(\pi-\young)\cos\young\right]\;
\label{eqn:hairCassie}
\end{equation}
may be thought of as the Cassie angle of the wetted section of the hairs. Eqn.~(\ref{eqn:freeEnergyZeroP3Db}) is equivalent in form to the free energy of the two-dimensional system (\ref{eqn:freeEnergyZeroP}), but with the replacements $\gamma\to\gamma D_{Y}$ (and thus $w\to\sqrt{\gamma D_{y}/K}L$) and $\young\to\cassie$. Hence the two-dimensional Eqns.~(\ref{eqn:stabilityYoung},\ref{eqn:selfConsistencyBaseYoung}-\ref{eqn:selfConsistencyJoinYoung}), with these replacements, may be used to describe the configuration of the hairs in the three-dimensional model.\\

A different modification is, however, needed to the depinning condition (\ref{eqn:depin}). If the interface were to depin from the tips of the hairs and descend a distance $-\delta\tilde{L}$, then the hitherto dry part of the hairs would be wetted, and the interface between hairs would be destroyed, leading to a free energy change.
\begin{equation}
\delta\mathcal{F}=\left\{\gamma\cos\young2a\young+\gamma(D_{y}-2a\sin\young)-\gamma D_{y}\cos\left[\alpha-\psi(L)\right]\right\}\delta\tilde{L}\;,
\end{equation}
 from which it may be seen that the condition for pinning of the interface on the tips of the hairs, $\delta\mathcal{F}/\delta{L}<0$, is given by (\ref{eqn:depin}), but with $\young$ replaced by an `antiCassie' angle~\cite{BicoTordeux}
\begin{equation}
\cos\anticassie=1-2\frac{a}{D_{y}}\left[\sin\young-\young\cos\young\right]\;.     
\label{eqn:hairantiCassie}
\end{equation}\\
Using this mapping, we show the regions of phase space where the singlet and doublet partially suspended states are stable or metastable in Fig~\ref{fig:phaseDiagrams3D}. The results presented are for vertical hairs, $\Omega=0$. Comparing to the equivalent two-dimensional plot, Fig.~\ref{fig:phaseDiagrams}, we see that the three dimensional system is much more successful than the two dimensional one in producing partially suspended states on hydrophilic materials. In particular, for doublets, the unzipping transition is shifted from $\young=90^{\circ}$ to $\cassie=90^{\circ}$ corresponding, for the parameters used here, to $\young=57^{\circ}$. Although the model predicts $\mathbb{U}$ states, we expect these to be unstable to $\mathbb{C}$, as the hydrophilic base substrate will promote spreading around the hairs. A higher $w$ is needed to stabilise the partially suspended states, and there is no critical point where the $\mathbb{P}_{1}\leftrightarrow \mathbb{F}$ transition becomes continuous.\\

As the interface bends to follow the profile of the hairs in the $x$ direction, it must have opposite curvature along $y$ to preserve the condition that the mean curvature of an equilibrium interface is zero. For small $D_{y}$, this curvature does not induce significant distortion of the interface but, as the spacing between the hairs increases, the approximation of two-dimensional hair profiles will become inappropriate. Moreover, the assumption that the hairs are constrained to bend only in the $x-z$ plane is artificial. Indeed, cylinders sitting on an interface are unstable to lateral perturbations~\cite{Crisp}, with distortions of the interface causing them to cluster. More work is needed to understand the interplay between the tendency to cluster, the curvature energy of the interface and the elastic energy of the hairs.\\

We now give estimates to relate the results to physical systems. The dimensions of the hairy surface of the Lady's Mantle are documented~\cite{MockForsterEtAl}, while the stiffness of the hairs is still not known. We use our theory to predict the elastic modulus that would allow the drop to be supported. We have noted that $w\sim w_{0}$ corresponds to the regime where the drop may be supported by hydrophilic hairs, giving
\begin{equation}
\sqrt{\frac{L}{D_{x}}}\sim\sqrt{\frac{\gamma D_{y}}{K}}L\;.
\end{equation}
Using measurements from the Lady's Mantle~\cite{MockForsterEtAl}, of $L\sim 1\;\mathrm{mm}$, $D_{x}=D_{y}\sim 0.5\;\mathrm{\mu m}$ and $a = 10\;\mathrm{\mu m}$, taking the surface tension $\gamma\sim 0.07\;\mathrm{Nm^{-1}}$ for water, and using the formula for $K=\tfrac{1}{4}\pi E a^{4}$ for a ~\cite{LandauLifschitz} cylindrical rod, we derive the suitable Young's modulus to be
\begin{equation}
E\sim\frac{\gamma D^{2} L}{a^{4}}\sim100\;\mathrm{MPa}\;
\end{equation}
which is a feasible value.\\

As a second example, we consider elastic posts, created articially by Mock et al~\cite{MockForsterEtAl}. These had dimensions $a\sim100\;\mathrm{\mu m}$, $d\sim500\;\mathrm{\mu m}$ and $L\sim500\;\mathrm{\mu m}$, and were made of HEMA with $E\sim0.5\;\mathrm{MPa}$~\cite{Krakovsky}. The posts were square in cross-section, hence $K=\tfrac{1}{12} E a^{4}$~\cite{LandauLifschitz}. Again taking $\gamma\sim 0.07\;\mathrm{Nm^{-1}}$, gives $w\sim0.5$, which is comparable to $w_{0}\sim1$. Unfortunately Mock et al could not obtain conclusive results for the interface position because of swelling of the posts.

\section{Discussion}

We have developed an elastocapillary model to quantify the wetting behaviour of a surface patterned 
with a regular array of elastic hairs. Our primary motivation was to identify superhydrophobic 
states asking, in particular, whether these could exist on hydrophilic hairs.\\

We began by working in two dimensions, considering a sufficiently large drop that interface 
curvature can be neglected, supported by a line of hairs or, equivalently, an array of parallel 
elastic sheets. This geometry allows an analytical solution. We identified three different states 
where the interface remains suspended above the base substrate. In addition to the usual 
Cassie-Baxter configuration, stable for rigid hairs, two partially suspended states can occur, the 
singlet ($\mathbb{P}_{1}$), where all hairs bend in the same direction, and the doublet 
($\mathbb{P}_{2}$) where neighbouring hairs bend in opposite directions. The partially suspended 
singlet can remain (meta)stable 
for both hydrophobic and hydrophilic hairs, for the dimensionless parameter $w=\sqrt{\gamma/K}L$ 
within a suitable range. Simulations showed that a drop placed gently on the surface is much more 
likely to fall into the state $\mathbb{P}_{1}$ if the hairs are rooted to the surface at a slight 
incline to the vertical.\\

The analysis naturally extends to a three dimensional system of hairs, given the restrictions that 
the hairs bend only in one plane, and the curvature of the interface perpendicular to this plane is 
minimal. The spacing of the hairs in the third dimension results in the Young angle being replaced 
by a Cassie angle with the consequence that, in contrast to two dimensions, hydrophilic hairs can 
support both singlet and doublet partially suspended states. The mechanism that stabilises these 
states is the propensity of the hairs to bind to the interface together with their resistance to 
bending. This is in contrast to that proposed in~\cite{OttenHerminghaus}, where the hairs pierce 
the surface. \\

The calculations reported were checked using both lattice Boltzmann and surface evolver 
calculations and, in future work, we aim to use the simulation approaches to obtain a more complete 
picture of what happens in three dimensions. It is important to understand the effect of 
instabilities which may cause the hairs to cluster: even in the rather artificial geometry 
considered in Sec.~\ref{section:3Dmodel}, simulations show a tendency for hairs to bunch together 
in the $y$ direction, and for more general hair configurations, we might expect more complicated 
clustering behaviour. Even the question of whether clustering bolsters~\cite{OttenHerminghaus} or 
hinders~\cite{ThorpeCrispD,Bush} superhydrophobicity still needs to be resolved.\\

Another question to consider is the edge effects arising from a finite size drop. For example, in 
the case of Lady's Mantle, the hair spacing can be of the order of the drop size. We speculate that 
hairs will most easily lie in the interface at the edges of the drop, and that this may help the 
system to form partially suspended states. It will also be interesting to investigate the dynamics 
of drops on hairy surfaces to ask, for example, whether the flexibility increases or decreases 
resistance to motion, and whether singlet states may provide a switchable anisotropy of the 
surface.

\section{Acknowledgements}

We thank A. Alexeev, A. C. Balazs, S. A. Brewer, H. Kusumaatmaja, G. McHale and N. J. Shirtcliffe for helpful discussions, N. R. Bernardino for providing a draft of Ref.~\cite{Bernardino} prior to publication, and L. Moevius for helpful comments on the manuscript.

\appendix
\section{Critical point for the partially suspended singlet at neutral wetting}

We already know that, for $\Omega=0$, the critical point lies on $\young=90^{\circ}$, and we now derive an analytical expression for the critical value of $w$. $\young=90^{\circ},s_{0}=1$ is always a solution to the equations for equilibrium
(\ref{eqn:selfConsistencyBaseYoung}-\ref{eqn:selfConsistencyAngleYoung}). By differentiating these equations with respect to the contact parameter $s_{0}$, we can determine whether the equilibrium is stable or unstable, using the criterion (\ref{eqn:stabilityYoung}). The derivatives of Eqns.~(\ref{eqn:selfConsistencyJoinYoung}) and (\ref{eqn:selfConsistencyAngleYoung}), are
\begin{align}
\frac{d\young}{ds_{0}}\Bigg\vert_{s_{0}=L}=\frac{d\psi(L)}{ds_{0}}\Bigg\vert_{s_{0}=L}-\frac{d\alpha}{ds_{0}}\Bigg\vert_{s_{0}=L}\;,  \label{eqn:diffSelfConsistencyJoinYoung}          \\
\frac{d\young}{ds_{0}}\Bigg\vert_{s_{0}=L}=\frac{d\alpha}{ds_{0}}\Bigg\vert_{s_{0}=L}\left(\lambda^{-2}LD-1\right)\;.   
\label{eqn:diffSelfConsistencyAngleYoung} 
\end{align}
In deriving (\ref{eqn:diffSelfConsistencyAngleYoung}) we use the results that $\tfrac{\partial}{\partial x}\tfrac{\cn_{m}x}{\dn_{m}x}=0$ and $\tfrac{\partial}{\partial m}\tfrac{\cn_{m}x}{\dn_{m}x}=0$ at  $x=0$ (see e.g.~\cite{Wolfram}). The terms of the derivative of the remaining equation of equilibrium (\ref{eqn:selfConsistencyBaseYoung}) cancel at $s_{0}=L$, so the second derivative is taken to give
\begin{equation}
\frac{d\psi(L)}{ds_{0}}\Bigg\vert_{s_{0}=L}^{2}=\lambda^{-2}L^{2}\left(\frac{d^{2}\young}{ds_{0}^{2}}\Bigg\vert_{s_{0}=L}-\frac{d^{2}\psi(L)}{ds_{0}^{2}}\Bigg\vert_{s_{0}=L}+\frac{d^{2}\alpha}{ds_{0}^{2}}\Bigg\vert_{s_{0}=L}\right)\;. 
\label{eqn:diffSelfConsistencyBaseYoung} 
\end{equation}
To eliminate $\tfrac{d^{2}\young}{ds_{0}^{2}}$ and $\tfrac{d^{2}\alpha}{ds_{0}^{2}}$, we take the second derivative of (\ref{eqn:selfConsistencyJoinYoung}) giving
\begin{equation}
\left(\frac{d\young}{ds_{0}}\Bigg\vert_{s_{0}=L}-\frac{d\alpha}{ds_{0}}\Bigg\vert_{s_{0}=L}\right)^{2}+2\frac{d^{2}\young}{ds_{0}^{2}}\Bigg\vert_{s_{0}=L}+2\lambda^{-2}=\frac{d\young(L)}{ds_{0}}\Bigg\vert_{s_{0}=L}^{2}+2\left(\frac{d^{2}\young}{ds_{0}^{2}}\Bigg\vert_{s_{0}=L}+\frac{d^{2}\alpha}{ds_{0}^{2}}\Bigg\vert_{s_{0}=L}\right)\;.         \label{eqn:secondDerivativeJoin}
\end{equation}
Solving (\ref{eqn:diffSelfConsistencyBaseYoung}-\ref{eqn:secondDerivativeJoin}) yields
\begin{equation}
\frac{d\young}{ds_{0}}\Bigg\vert_{s_{0}=L}=D^{-1}-\lambda^{-2}L\;.     \label{eqn:thetaPrime}
\end{equation}
Thus, the condition for stability (\ref{eqn:stabilityYoung}) becomes
\begin{equation}
\lambda<\lambda_{0}:=\sqrt{LD}\;,          \label{eqn:criticalPointLam}
\end{equation}
or, in terms of the dimensionless parameter $w$,
\begin{equation}
w>w_{0}:=\sqrt{\frac{L}{D}}\;.          \label{eqn:criticalPoint}
\end{equation}

\end{document}